\documentclass[a4paper,12pt]{extarticle}

\usepackage[margin=1.0in]{geometry}
\usepackage[english]{babel}
\usepackage{amsmath}
\usepackage{parskip}
\usepackage{wrapfig}
\usepackage{graphicx}
\usepackage{tikz-cd}
\usepackage{lipsum}
\usepackage{adjustbox}
\usepackage{amssymb}
\usepackage{tikz}
\usepackage{hyperref}
\usepackage{amsmath}
\usepackage{amsfonts}
\usepackage{amssymb}
\usepackage{amsthm}
\usepackage{amsbsy}
\usepackage{bm}
\usepackage{mathrsfs}
\usepackage{slashed}
\usepackage{graphicx}
\usepackage{fancyhdr}
\usepackage{cite}
\usepackage{calc}

\title{\textbf{Type IIB superstring vertex operator from the -8 picture}}
\author{Lucas N.S. Martins\thanks{lucas\underline{\hspace{.06in}}nmartins@hotmail.com}}

\date{}

\begin{document}

% ------------- Making title -------------

\begin{titlepage}
\pagenumbering{gobble}
\maketitle
\vspace{-1.0cm}
\begin{center}
\small{\textit{Instituto de F\'{i}sica Te\'{o}rica, Universidade Estadual Paulista, \\
Rua Dr. Bento Teobaldo Ferraz 271, \\
Bloco II -- Barra Funda, \\
CEP: 01140-070 -- S\~{a}o Paulo, Brasil.}}
\end{center}
\vspace{.1cm}

% ------------- Making abstract -------------

\begin{abstract}
A new procedure was recently proposed for constructing massless Type IIB vertex operators in the pure spinor formalism. Instead of expressing these closed string vertex operators as left-right products of open string vertex operators, they were instead constructed from the complex N=2 d=10 superfield whose lowest real and imaginary components are the dilaton and Ramond-Ramond axion. These Type IIB vertex operators take a simple form in the -8 picture and are related to the usual vertex operators in the zero picture by acting with picture-raising operators. In this paper, we compute explicitly this picture-raising procedure and confirm this proposal in a flat background. Work is in progress on confirming this proposal in an $AdS_5\times S^5$ background.
\end{abstract}
\end{titlepage}

\pagenumbering{arabic}

\section{Introduction}

In 2000 a new formalism for superstring theory was proposed, namely the pure spinor formalism \cite{Berkovits:2000fe}. This new formalism has the advantage of being super Poincaré covariant, solving various problems of previous formalism \cite{Friedan:1985ge} \cite{Green:1980zg}, such as the proof of vanishing theorems \cite{Berkovits:2004px}, the action for the superstring in the presence of Ramond-Ramond flux \cite{Berkovits:2000fe}\cite{Berkovits:2008ga}\cite{Berkovits:2012ps} and the computation of certain  multi-loop scattering amplitudes \cite{Gomez:2013sla}.

Recently the pure spinor formalism was applied to the small radius limit of Type IIB superstrings in $AdS_{5}\times S^{5}$ \cite{Berkovits:2019ulm}. It was argued that the Feynman diagrams of $\mathcal{N}=4$ $d=4$ Super-Yang-Mills for large $N$ can be reproduced by a given string amplitude prescription in the $AdS_5\times S^5$ background. A crucial ingredient for this amplitude prescription was a conjecture that a half-BPS vertex operator, in the "-8 picture", is given by

\begin{equation}
V_{-8}=(\lambda_L\eta\lambda_R)A_{n}\prod_{A=1}^{8}\theta^{A}\delta(Q(\theta^{A}))    
\end{equation}

where $A$ is a bosonic function that is completely determined by the dimension and $R$ charge of the half-BPS state, the eight fermionic variables $\theta^{A}$ are related to the eight supersymmetries that do not annihilate the half-BPS state, $\lambda_{L}^{\alpha}$ and $\lambda_{R}^{\alpha}$ are the left and right pure spinors, $\eta_{\alpha\beta}$ is the bi-spinor that sets the direction of the Ramond-Ramond flux, $Q$ is the sum of the left and right-moving BRST operators, and $\alpha$ and $\beta$ go from $1$ to $16$ and are the Majorana-Weyl spinorial indices for ten dimensions. The conjecture states that if we hit with an appropriate "picture raising" operator, we obtain the vertex operator in "zero picture" of the form $V=\lambda_{L}^{\alpha}\lambda_{R}^{\beta}A_{\alpha\beta}$, without delta-functions.

In a subsequent paper \cite{Berkovits:2019rwq}, this conjecture was extended to all Type IIB backgrounds including the flat background. Since the vertex operators for the flat background are known, this is a good place for a test of the conjecture which will be the content of this paper. Here, the "-8 picture" vertex operator for the whole supergravity multiplet in a flat background will be constructed and the "0 picture" vertex operator will be computed from this procedure of picture raising. The result obtained here is the expected one written in terms of a light-cone superfield whose lowest real and imaginary component are the dilaton and the Ramond-Ramond axion \cite{Berkovits:2019rwq}. This superfield is known to describe light-cone Type IIB supergravity \cite{Green:1983hw}.

For the case of the flat background, in coordinates where  $k_{+}=k_{9}+k_{0}$ is the only non-zero component for the momentum, the "-8 picture" vertex operator is given by

\begin{equation}
V_{-8}=\frac{(\lambda_{L}\gamma^{-}\lambda_{R})}{(4ik_+)^4} \bar\Phi(\bar y,\theta_{+})\prod_{a=1}^{8}\delta (Q(\theta^{a}_{+})) \end{equation}

where $\lambda^{\alpha}_{L}$ and $\lambda^{\beta}_{R}$ are the Type IIB pure spinors, $\theta^{a}_{+}\equiv (\gamma^{+}(\theta_{L}+i\theta_{R}))^{a}$ are the eight fermionic variables that span the light-cone supergravity multiplet, and $\bar\Phi$ is the superfield of \cite{Green:1983hw}. In \cite{Berkovits:2019rwq} a reduced version of this vertex operator was proposed, where just the scalar states of the supergravity multiplet were included. Here we will generalize to the whole multiplet by covariatizing with respect to the supersymmetries (see equation (29) and (30)).

As suggested in \cite{Berkovits:2019rwq}\cite{Berkovits:2019ulm}, the zero picture operator can be obtained from the $-8$ picture by acting with eight picture-raising operators, which gives rise to the following equation for the zero picture vertex operator $V$

\begin{equation}
V=\left(\prod_{a=1}^{8}Q(\xi^{(a)})\right)V_{-8}    
\end{equation}

where $Q(\xi^{(a)})$ are the picture raising operators. The result obtained here will be that 
$V=V_0+V_1+V_2+V_3+V_4$ with $V_n$ given by

\begin{equation}
V_{n}=\frac{(\bar\lambda_L\bar\sigma^{i_1j_1\dots i_nj_n}\bar\lambda_R)}{n!(32ik_+)^n}(\nabla_+\sigma^{i_1j_1}\nabla_+)\dots  (\nabla_+\sigma^{i_nj_n}\nabla_+)  \Phi(y,\theta_{-})
\end{equation}

This result agrees with what was expected in \cite{Berkovits:2019rwq}, where there the vertex operator was computed directly by solving the equations of BRST invariance in a particular gauge characterized by the decoupling of half the fermionic variables.

Since this conjecture was confirmed here for the flat background, we can try to use it to obtain vertex operators for the $AdS_{5}\times S^{5}$ background. Up to now, the explicit vertex operators in the zero picture for the $AdS_{5}\times S^{5}$ background are unknown and this new procedure might be our best hope to obtain them. Work is in progress on obtaining the half-BPS vertex operators using this conjecture.

In section 2 we review the supergravity vertex operator for a flat background and fix notation, where in subsection 2.1 we solve the pure spinor constraint in a particular way that will be useful for later, and in subsection 2.2 we review the construction of the $-8$ picture vertex operator, and extend to the whole supermultiplet. In section 3 we compute the vertex operator in the 0 picture by acting with 8 picture raising operators on the -8 picture vertex operator, where in subsection 3.1 we work out some initial steps in the computation, and in subsection 3.2, we use the method of induction to finally compute the vertex operator in the 0 picture. In section 4 we conclude. 

\section{Review}

In any type IIB supergravity background, a massless closed string vertex operator in unintegrated form in the pure spinor formalism can be written as \cite{Berkovits:2000yr}

\begin{equation}
V=\lambda_{L}^{\alpha }\lambda_{R}^{\beta} A_{\alpha\beta }(x,\theta)  
\end{equation}

where $\alpha=1$ to $16$ are the Weyl-Majorana $SO(1,9)$ indices. The coordinates $(x,\theta)\equiv (x^{\underline{m}},\theta^{\alpha}_{L},\theta_{R}^{\alpha})$, with $\underline{m}=0$ to $9$, parametrize the Type IIB superspace. The pair of Weyl spinors $(\lambda_{L}^{\alpha},\lambda_{R}^{\alpha})$ are the left and right pure spinors, constrained to satisfy $(\lambda_{L}\gamma^{\underline{m}}\lambda_{L})=(\lambda_{R}\gamma^{\underline{m}}\lambda_{R})=0$. The $(\gamma^{m}_{\alpha\beta},\gamma_{m}^{\alpha\beta})$ are the real symmetrical gamma matrices for $SO(1,9)$. The on-shell equations of motion comes from $QV=0$ and the gauge invariances comes from $V\cong V+Q\Lambda$, where $Q= \lambda^{\alpha}_{L}\nabla_{\alpha}^{L}+\lambda^{\alpha}_{R}\nabla_{\alpha}^{R}$. The supercovariant derivatives $(\nabla^{L}_{\alpha},\nabla^{R}_{\alpha})$ are

\begin{equation}
\nabla^{L}_{\alpha}=\frac{\partial}{\partial\theta_{L}^{\alpha}}+(\gamma^{\underline{m}}\theta_{L})_{\alpha}\frac{\partial}{\partial x^{\underline{m}}},\qquad\nabla^{R}_{\alpha}=\frac{\partial}{\partial\theta_{R}^{\alpha}}+(\gamma^{\underline{m}}\theta_{R})_{\alpha}\frac{\partial}{\partial x^{\underline{m}}}     
\end{equation}

In the light-cone frame, where $k_+=k_{0}+k_{9}$ is the only nonzero component for the momentum, there is a convenient gauge fix characterized by the decoupling of half of the fermionic coordinates \cite{Berkovits:2019rwq}. In this frame, the $SO(1,9)$ Weyl spinors $\chi^{\alpha}$ naturally decomposes into $SO(8)$ light-cone spinors

\begin{equation}
\chi^{a}=\left(\frac{(\gamma_{9}+\gamma_{0})}{2}\chi\right)^{a},\qquad \bar\chi^{\dot a} =  \left(\frac{(\gamma_{9}-\gamma_{0})}{2}\chi\right)^{\dot a} \end{equation}

where $a,\dot a=1$ to $8$ are the Weyl and anti-Weyl spinorial indices respective to the $SO(8)$ light-cone indices. As we are going to see next, the fermionic coordinates that decouples are the $\bar\theta_{L}^{\dot a}$ and $\bar\theta_{R}^{\dot a}$. 

The pure spinor constraints $(\lambda_{L}\gamma^{\underline{m}}\lambda_{L})=(\lambda_{R}\gamma^{\underline{m}}\lambda_{R})=0$ in $SO(8)$ notation becomes 

\begin{equation}
\lambda_L^a\lambda_L^a=\bar\lambda_L^{\dot a}\bar\lambda_L^{\dot a}=\lambda_R^a\lambda_R^a=\bar\lambda_R^{\dot a}\bar\lambda_R^{\dot a}=0,\qquad\lambda_{L}^a\sigma^m_{a\dot a}\bar\lambda_L^{\dot a}=   \lambda_{R}^a\sigma^m_{a\dot a}\bar\lambda_R^{\dot a}=0 
\end{equation}

where $m=1$ to $8$ and $\sigma^{m}_{a\dot a}$ are the $SO(8)$ real Pauli matrices.

The supercovariant derivatives in $SO(8)$ light-cone notation reduces to

\begin{equation}
\nabla_{a}^{L}= \frac{\partial}{\partial \theta_{L}^{a}}+ik_+\theta_{La},\quad \nabla_{a}^{R}= \frac{\partial}{\partial \theta_{R}^{a}}+ik_+\theta_{Ra},\quad \bar\nabla_{\dot a}^{L}= \frac{\partial}{\partial \bar\theta_{L}^{\dot a}},\quad \bar\nabla_{\dot a}^{R}= \frac{\partial}{\partial \bar\theta_{R}^{\dot a}}    
\end{equation}

which allow us to gauge fix $A_{ab} =A_{\dot a b}=A_{ a\dot b}=0$, getting the vertex operator

\begin{equation}
V=\bar\lambda_{L}^{\dot a}\bar\lambda_{R}^{\dot b}A_{\dot a\dot b}(k_+,\theta_{L},\theta_{R})e^{ik_+x^+}     
\end{equation}

where BRST invariance implies 

\begin{equation}
\frac{\partial A_{\dot a\dot b}}{\partial\bar\theta_L^{\dot c}}= \frac{\partial A_{\dot a\dot b}}{\partial\bar\theta_R^{\dot c}} =0,  \qquad \nabla^{L}_{a}A_{\dot b \dot c} = \frac{1}{8}\sigma^{m}_{a\dot b}\sigma_{m}^{d\dot e}\nabla^L_{d}A_{\dot e \dot c},\qquad \nabla^{R}_{a}A_{\dot b \dot c} = \frac{1}{8}\sigma^{m}_{a\dot c}\sigma_{m}^{d\dot e}\nabla^R_{d}A_{\dot b \dot e} 
\end{equation}

Note that $\bar\theta_L^{\dot a}$ and $\bar\theta_R^{\dot a}$ decouples. Instead of solving these equations by brute force we are going to construct a -8 picture vertex operator. Hitting 8 picture raising operators gives the solution to the above equations.

It will be convenient for us to organize the $SO(8)$ light-cone spinors as

\begin{equation}
\theta_{\pm}^{a}=\theta_{L}^{a}\pm i \theta_{R}^{a},\quad \nabla_{\pm}^{a}= \nabla_{L}^{a}\pm i \nabla_{R}^{a},\quad \lambda_{\pm}^{a}=\lambda_{L}^{a}\pm i \lambda_{R}^{a},\quad \bar\lambda_{\pm}^{\dot a}=\bar \lambda_{L}^{a}\pm i \bar\lambda_{R}^{\dot a}
\end{equation} 

The solution to the equations in (10), in terms of the variables above, was obtained in \cite{Berkovits:2019rwq} and is given by

\begin{equation}
V=V_0+V_1+V_2+V_3+V_4 
\end{equation}

with

\begin{equation}
V_n=\frac{(\bar\lambda_L\sigma^{i_1j_1\dots i_nj_n}\bar\lambda_R)}{n!(32ik_+)^n}(\nabla_+\sigma^{i_1j_1}\nabla_+)\dots  (\nabla_+\sigma^{i_nj_n}\nabla_+) \Phi(y,\theta_-)  
\end{equation}

where $\Phi(y,\theta_{-})$ being the superfield of \cite{Green:1983hw} obeying the constraints

\begin{equation}
\frac{\partial}{\partial x^{\underline{m}}}\Phi = i\delta_{\underline{m}}^{+}k_{+}\Phi,\qquad\nabla_{-}^a\Phi = 0,\qquad (\nabla_{+})^{4}_{abcd}\Phi=\frac{1}{4!}\varepsilon_{abcdefgh}(\nabla_{-})^{4}_{efgh}\bar\Phi    
\end{equation}

with $\bar\Phi$ being the complex conjugate of $\Phi$. The first and second constraint can be solved by $\Phi(y,\theta_-)=\Phi(k_+,\theta_{-})e^{ik_+y}$, with $y=x_++ 2i\theta_+\theta_-$. The third constraint is a reality condition that reduces the $2^8$ complex components to $2^8$ real components, the right number of components to describe the Type IIB supergravity multiplet. In what follows we are going to obtain the same vertex operator by the procedure proposed in \cite{Berkovits:2019ulm}\cite{Berkovits:2019rwq} which involves the picture raising of the $-8$ picture vertex operator.

\subsection{Pure Spinor Constraints}

The ten dimensional pure spinor constraints in terms of the variables $\lambda_{\pm}^{a},\,\bar\lambda_{\pm}^{\dot a}$ are

$$
\lambda^{a}_{+}\lambda^{a}_{+}+\lambda^{a}_{-} \lambda^{a}_{-}=0,\qquad\bar\lambda^{\dot a}_{+}\bar\lambda^{\dot a}_{+}+\bar\lambda^{\dot a}_{-}\bar\lambda^{\dot a}_{-}=0,\qquad\lambda_{+}^{a}\lambda_{-}^{a}=0,\qquad\bar\lambda_{+}^{\dot a}\bar\lambda_{-}^{\dot a}=0
$$

\begin{equation}
\lambda_{+}^{a}\sigma_{a\dot a}^{m}\bar\lambda_{+}^{\dot a}+\lambda_{-}^{a}\sigma_{a\dot a}^{m}\bar\lambda_{-}^{\dot a}=0,\qquad \lambda_{+}^{a}\sigma_{a\dot a}^{m}\bar\lambda_{-}^{\dot a}+\lambda_{-}^{a}\sigma_{a\dot a}^{m}\bar\lambda_{+}^{\dot a}=0
\end{equation}

Using the $SO(8)$ Pauli matrices identity

\begin{equation}
\delta_{ab}\delta_{\dot a\dot b}=\sigma^{m}_{a\dot b}\sigma^{m}_{\dot a b}-\frac{1}{4}\sigma^{mn}_{ab}\bar\sigma^{mn}_{\dot a\dot b}
\end{equation}

together with the pure spinor constraints, we can write

\begin{equation}
(\sigma^{m}\lambda_{+})_{\dot a}(\sigma^{m}\lambda_{-})_{\dot b} = \frac{1}{4}(\lambda_{+}\sigma^{mn}\lambda_{-}) \sigma^{mn}_{\dot a \dot b}
\end{equation}

With this formula above it is possible to solve the pure spinor constraint by expressing $\lambda_{-}^{a}$ in terms of $\lambda_{+}^a$ and $\bar\lambda_{\pm}^{\dot a}$ as follows

$$
\lambda_{-}^{a}=\frac{\bar\lambda_{+}^{\dot a}\bar\lambda_{+}^{\dot a}}{(\bar\lambda_{+}\bar\lambda_{+})}\lambda_{-}^{a} = \frac{\bar\lambda_{+}^{\dot a}}{(\bar\lambda_{+}\bar\lambda_{+})}\left((\sigma^{m}\bar\lambda_{+})^{a}(\bar \sigma^{m}\lambda_{-})^{\dot a}-\frac{1}{4}(\sigma^{mn}\bar\lambda_{+})^{\dot a}(\sigma^{mn}\lambda_{-})^{ a}\right) 
$$

\begin{equation}
=(\sigma^{m}\bar\lambda_{+})^{a}\frac{(\bar\lambda_{+}\sigma^{m}\lambda_{-})}{(\bar\lambda_{+}\bar\lambda_{+})}= - (\sigma^{m}\bar\lambda_{+})^{a}\frac{(\lambda_{+}\sigma^{m}\bar\lambda_{-})}{(\bar\lambda_{+}\bar\lambda_{+})}=  \frac{1}{4}\frac{(\bar\lambda_{+}\bar\sigma^{mn}\bar\lambda_{-})}{(\bar\lambda_{+}\bar\lambda_{+})}(\sigma^{mn}\lambda_{+})^{a}
\end{equation}

This means that if $(\bar\lambda_{+}\bar\lambda_{+})\neq 0$ one can solve the pure spinor constraints by fixing

\begin{equation}
\lambda_{-}^{a}\equiv\frac{(\bar\lambda_{+}\sigma^{mn}\bar\lambda_{-})}{4(\bar\lambda_{+}\bar\lambda_{+})}(\sigma^{mn}\lambda_{+})^{a}=\frac{(\bar\lambda_{R}\sigma^{mn}\bar\lambda_{L})}{4(\bar\lambda_{L}\bar\lambda_{R})}(\sigma^{mn}\lambda_{+})^{a}
\end{equation}

In other words, in the patch where $(\bar\lambda_{L}\bar\lambda_{R})\neq 0$, a pure spinor can be parametrized by 8 components of $\lambda_{+}^a$, and 14 components of $\bar \lambda_{L}^{\dot a}$ and $\bar\lambda_{R}^{\dot a}$ which are constrained to satisfy $(\bar\lambda_{L}\bar\lambda_{L})=(\bar\lambda_{R}\bar\lambda_{R})=0$. Note that we have in total 22 components, the right number of components for a pair of pure spinors. It is important in what follows that all these 8 $\lambda_+^a$ components are independent. 

\subsection{The -8 picture vertex operator}

For the parametrization of the pure spinor space above, we can treat the system $(w_{a}^{+},\lambda^{a}_{+})$ as a free system. Note that this free system is not a conventional $\beta$-$\gamma$ system. This is so because $\lambda_{+}^{a}$ mixes holomorphic and antiholomorphic fields. Nonetheless, inspired by the Friedan-Martinec-Shenker bosonization formulas \cite{Friedan:1985ge}\cite{Witten:2012ga}

$$
\lambda^{a}_{+}\cong \eta^{(a)}e^{\phi^{(a)}},\quad w_{a}^{+}\cong e^{-\phi^{(a)}}\partial\xi^{(a)}
$$

\begin{equation}
\delta(\lambda^{a}_{+})\cong e^{-\phi^{(a)}},\quad (\lambda^{a}_{+})^{-1}\cong \xi^{(a)}e^{-\phi^{(a)}}
\end{equation}

we can postulate the existence of fields $\xi^{(a)}$ that satisfy  $\xi^{(a)}\delta(\lambda^{a}_{+})=(\lambda^{a}_{+})^{-1}$. In fact it is possible to write a formal expression for $\xi^{(a)}$ in terms of $\lambda_{+}^{a}$:

\begin{equation}
\xi^{(a)}=\int \frac{1}{\tau}e^{-\tau \frac{\partial}{\partial \lambda_{+}^{a}}}d\tau
\end{equation}

Representing the delta function $\delta(\lambda_{+}^{a})$ in integral form

\begin{equation}
\delta(\lambda_{+}^{a})=\int\frac{d\omega}{2\pi} e^{i\omega\lambda^{a}_{+}}    
\end{equation}

we can see that

$$
\xi^{(a)}\delta(\lambda_{+}^{a})=\int d\tau\int \frac{d\omega}{2\pi} \frac{1}{\tau}e^{-\tau \frac{\partial}{\partial \lambda_{+}^{a}}} e^{i\omega\lambda^{a}_{+}}=\int d\tau\int \frac{d\omega}{2\pi} \frac{1}{\tau} e^{i\omega(\lambda^{a}_{+}-\tau)} =
$$

\begin{equation}
=\int d\tau \frac{1}{\tau} \delta(\lambda^{a}_{+}-\tau) = \frac{1}{\lambda^{a}_{+}}
\end{equation}

as desired.

Insisting in this analogy with bosonization, the picture number might be defined as the conserved charge 

\begin{equation}
P=-\sum_{a=1}^{8}\frac{1}{2\pi i}\oint_{C} \frac{d \lambda^{a}_{+}}{\lambda^{a}_{+}}\cong-\frac{1}{2\pi i}\sum_{a=1}^{8}\left(\oint_{C}\partial\phi^{(a)}+\oint_{C}\eta^{(a)}\xi^{(a)}\right)    
\end{equation} 

This motivate the assignment of a picture number $-1$ for $\delta(\lambda^{a}_{+})$ and $+1$ for $\xi^{(a)}$. In fact we can evaluate explicitly

\begin{equation}
\frac{1}{2\pi i}\oint_{C} \frac{d \lambda^{a}_{+}}{\lambda^{a}_{+}}\delta(\lambda_{+}^{a})= \delta(\lambda_{+}^{a})    
\end{equation}

where the contour $C$ goes around $\delta(\lambda_{+}^{a})$. This is because the delta function restrict $\lambda_{+}^{a}(z,\bar z)$ to map the point where the $\delta(\lambda^{a}_{+})$ is inserted in the worldsheet to the point $\lambda_{+}^{a}=0$, which implies that the contour in the worldsheet is mapped into a contour around $\lambda_{+}^{a}=0$. 

Using both $\delta(\lambda^{a}_{+})$ and $\xi^{(a)}$ it is possible to construct two "BRST exact" operators given formally by

\begin{equation}
X^{(a)}=Q(\xi^{(a)})\quad\text{and}\quad Y^{(a)}=\theta_{+}^{a}\delta(\lambda_+^a)\cong Q(\xi^{(a)} e^{-2\phi^{(a)}}\theta^{(a)}_{+}\partial\theta^{(a)}_{+}) \end{equation} 

They will be called, respectively, the picture raising operator and picture lowering operator. Note that $X^{(a)}$ has picture +1 and $Y^{(a)}$ has picture -1, and both fields has zero conformal weight.

The $\bar\lambda_{\pm}^{\dot a}$ components of the pure spinors are independent of $\lambda_{+}^{a}$, so they commute with $\xi^{a}$ and $\delta(\lambda_{+}^{a})$. However, the $\lambda_{-}^{a}$ components does depend on $\lambda_{+}^{a}$ through (20), implying

\begin{equation}
\lambda_{-}^{b}\prod_{a=1}^{8}\theta_{+}^{a}\delta(\lambda_{+}^{a})=0    
\end{equation}

i.e. they are annihilated by all the 8 picture lowering operators. This simplify the computation of the BRST cohomology in the $-8$ picture.

A scalar state $|A\rangle$ that is annihilated by the supercharges $q_{+}^{a}$ and $\bar q_{\pm}^{\dot a}$ have as a -8 picture vertex operator

\begin{equation}
V_{-8}^{|A\rangle}=(\bar\lambda_{L}\bar\lambda_{R})e^{ik_{+}x^{+}}\prod_{a=1}^{8}\theta_{+}^{a}\delta(\lambda^{a}_{+})    
\end{equation}

Acting with the $q_{-}^{a}$ we can generate the full supergravity multiplet. The -$8$ picture massless vertex operator $V_{-8}$ obtained by this super covariantization is

\begin{equation}
V_{-8}=\frac{(\bar \lambda_{L}\bar\lambda_{R})}{(4ik_+)^8}\left(\prod_{a}^{8}\delta (\lambda^{a}_+)\nabla^{a}_+\right) \Phi(y,\theta_{-})
\end{equation}

where $\Phi(y,\theta_{-})$ is the complex superfield satisfying the constraints of (15). We can easily verify that this vertex operator is BRST invariant and it has the right number of components to describe the Type IIB supergravity multiplet. Note that this vertex operator simplify if we write in terms of the complex conjugate $\bar\Phi(\bar y,\theta_{+})$

\begin{equation}
V_{-8}=\frac{(\bar \lambda_{L}\bar\lambda_{R})}{(4ik_+)^4}\bar\Phi(\bar y,\theta_+)\prod_{a}^{8}\delta (\lambda^{a}_+)
\end{equation}

In our conventions $[\nabla_{+}^{a},\nabla_{-}^{b}]=4ik_+\delta^{ab}$. The powers of $4ik_+$ are fixed by requiring that the first component of $\Phi(y,\theta_{+})$ is a complex field where the real part is the dilaton and the imaginary part is the Ramond-Ramond axion.

\section{The zero picture vertex operator from the -8 picture}

The zero picture vertex operator $V$ can be obtained by hitting 8 picture raising operators in $V_{-8}$

\begin{equation}
V=\left(\prod_{a=1}^{8} Q(\xi_{a})\right)\left(\prod_{a=1}^{8}\delta(\lambda_{+}^{a})\nabla_{+}^{a}\right)\frac{(\bar\lambda_{L}\bar\lambda_{R})}{(4ik_+)^8}\Phi(y,\theta_{-})
\end{equation}

where $\Phi(y,\theta_-)\equiv\Phi(k_+,\theta_-)e^{ik_+y}$. By the nilpotence of the BRST charge and the BRST invariance of $V_{-8}$ we can write

\begin{equation}
V= Q\left(\xi^{(8)}\dots Q\left(\xi^{(1)}\left(\prod_{a=1}^{8}\delta(\lambda_{+}^{a})\nabla_{+}^{a}\right)\frac{(\bar\lambda_{L}\bar\lambda_{R})}{(4ik_+)^8}\Phi(y,\theta_{-})\right)\dots\right)
\end{equation}    

Each $Q(\xi^{(a)})$ split into two operators. One operator that absorbs one $\nabla_+^a$ and another that adds a new one. This imply that each term produced by the action of all the 8 picture changing operators in $V_{-8}$ will be given by a sequence of these two operators. Explicitly, at the $n$-th picture raising, the operators, denoted by $d$ and $u$, will be:

$$d(\delta(\lambda^{(b)}_{+})\nabla_{+}^{(b)}\nabla_{+}^{a_{1}}\dots\nabla_{+}^{a_n})=  \frac{(4ik_+)}{\lambda_{+}^{b}}\lambda_{+}^{[b}\nabla_{+}^{a_1}\dots\nabla_{+}^{a_n]} 
$$

\begin{equation}
u(\delta(\lambda^{(b)}_{+})\nabla_{+}^{(b)}\nabla_{+}^{a_{1}}\dots\nabla_{+}^{a_n})=(\lambda_{-}\nabla_{+}) \frac{\nabla_{+}^{b}}{\lambda_{+}^{b}}\nabla_{+}^{a_{1}}\dots\nabla_{+}^{a_n}
\end{equation}

It is convenient to represent diagrammatically these sequences of $u$ and $d$ as paths in a diagram constructed in such a way that the numbers of thetas, deltas and $Q(\xi)$ can be read immediately.

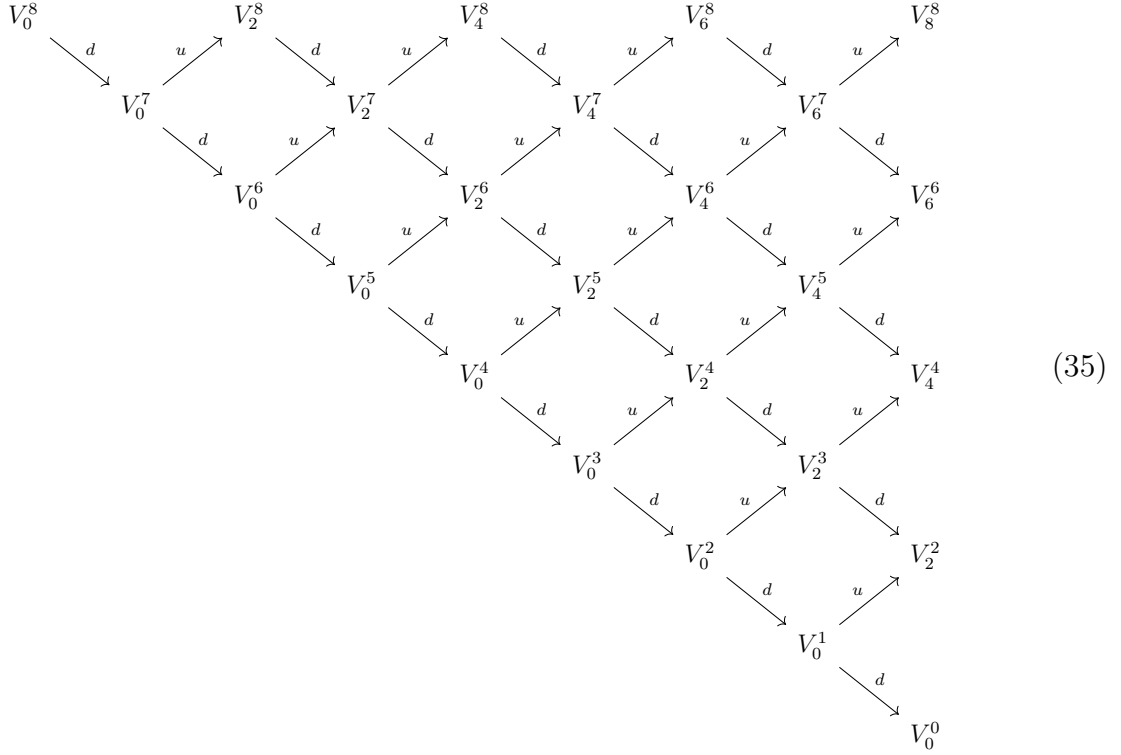
\begin{equation}
\begin{tikzpicture}[baseline= (a).base]
\node[scale=.8] (a) at (0,0){
\begin{tikzcd}[scale=0.6]
&V_{0}^8 \arrow{dr}{d} && V_{2}^8 \arrow{dr}{d} && V_{4}^8 \arrow{dr}{d} && V_{6}^8 \arrow{dr}{d} && V_{8}^8\\&
& V_{0}^7 \arrow{dr}{d}\arrow{ur}{u} && V_{2}^7 \arrow{dr}{d}\arrow{ur}{u} && V_{4}^7 \arrow{dr}{d}\arrow{ur}{u} && V_{6}^7 \arrow{dr}{d}\arrow{ur}{u}\\&
&& V_{0}^6 \arrow{dr}{d}\arrow{ur}{u} && V_{2}^6 \arrow{dr}{d}\arrow{ur}{u} && V_{4}^6 \arrow{dr}{d}\arrow{ur}{u} && V_{6}^6 \\&
&&& V_{0}^5\arrow{dr}{d}\arrow{ur}{u} && V_{2}^5 \arrow{dr}{d}\arrow{ur}{u} && V_{4}^5 \arrow{dr}{d}\arrow{ur}{u} \\&
&&&& V_{0}^4 \arrow{dr}{d}\arrow{ur}{u}  && V_{2}^4 \arrow{dr}{d}\arrow{ur}{u} && V_{4}^4\\& &&&&& V_{0}^3 \arrow{dr}{d}\arrow{ur}{u} && V_{2}^3 \arrow{dr}{d}\arrow{ur}{u} \\& &&&&&& V_{0}^2 \arrow{dr}{d}\arrow{ur}{u} && V_{2}^2 \\& &&&&&&& V_{0}^1 \arrow{dr}{d}\arrow{ur}{u} && \\& &&&&&&&& V_{0}^0 
\end{tikzcd}
};
\end{tikzpicture}
\end{equation}

A path following the arrows correspond to a given sequence of $u$ and $d$, and the terms $V_{n}^{m}$ in the diagram correspond to an expression with $m$ $\nabla_{+}$ at a given stage of a sequence that will contribute to the term with $n$ $\nabla_+$ at the end, namely the $V_{n}^{n}$ (this is related to the number of deltas). The arrows pointing down (up) represent the $d$ ($u$) transformation. Note that our 0 picture vertex operator $V$ will split into a sum of five terms $V_0^0+V_2^2+V_4^4+V_6^6+V_8^8$, with 0, 2, 4, 6 and 8 $\nabla_+$.

Using our parametrization for the solution of the pure spinor constraints, the $u$ operation reads

\begin{equation}
u(\delta(\lambda^{(b)}_{+})\nabla_{+}^{(b)}\nabla_{+}^{a_{1}}\dots\nabla_{+}^{a_n})= \frac{ (\bar\lambda_{L}\sigma^{mn}\bar\lambda_{R})}{4(\bar\lambda_{L}\bar\lambda_{R})}\frac{(\nabla_{+}\sigma_{mn}\lambda_{+})}{\lambda_{+}^{b}}\nabla_{+}^{b}\,\nabla_{+}^{a_{1}}\dots\nabla_{+}^{a_n}     
\end{equation}

where the dependence in $\lambda_{+}^{a}$ is explicit and terms that are non-singular at $\lambda_{+}^{a}=0$ can be easily separated from terms with poles.

The $V_{0}^{0}$ term is easy to be obtained since it has a unique path $d^{8}$. It is given by

\begin{equation}
V_{0}^0=(\bar\lambda_L\bar\lambda_R)\Phi(y,\theta_-)
\end{equation}

The $V_{2}^2$ has 7 paths and can be obtained by looking at the path $dud^{6}$ that produces

$$
\rightarrow_{d}\frac{(\bar\lambda_{L}\bar\lambda_{R})}{(4ik_+)^7}\nabla_{+}^{2}\delta(\lambda_{+}^{2})\dots\rightarrow_{u}\frac{(\bar\lambda_{L}\bar\lambda_{R})}{(4ik_+)^7}\left(\frac{ (\bar\lambda_{L}\sigma^{ij}\bar\lambda_{R})(\nabla_{+}\sigma^{ij}\lambda_{+})}{4(\bar\lambda_{L}\bar\lambda_{R})}\right)\frac{\nabla_{+}^{2}}{\lambda_{+}^{2}}\dots
$$

\begin{equation}
\rightarrow_{d^{6}}\frac{(\bar\lambda_{L}\sigma^{ij}\bar\lambda_{R})}{16ik_+}\sigma^{12}_{ij}\nabla_{+}^{1}\nabla_{+}^{2}\Phi(y,\theta_{-}) + \text{poles}
\end{equation}

Note that consistency requires that the 0 picture vertex operator $V$ should have no poles in $\lambda_{+}^a=0$. We are going to show in the next section that indeed all the poles cancel, but for now let us just drop them. Since this is the only path that has a chance of producing both $\nabla_{+}^{1}$ and $\nabla_{+}^{2}$, without poles, by $SO(8)$ symmetry we have

\begin{equation}
V_{2}^2=\frac{(\bar\lambda_{L}\sigma^{ij}\bar\lambda_{R})}{32ik_+} (\nabla_{+} \sigma_{ij}\nabla_{+})\Phi(y,\theta_{-})
\end{equation}

which agrees with \cite{Berkovits:2019rwq}. It is instructive to see how each path contribute to various components of the above expression. Denoting the term proportional to $\nabla^{a}_{+}\nabla^{b}_{+}$ by $[ab]$, we have

$$
dud^6:\,[12]; \qquad d^2ud^5:\,[13],[23]; \qquad d^3ud^4:\,[14],[24],[34];
$$

\begin{equation}
d^4ud^3:\,[15],[25],[35],[45];\qquad d^{n-1}ud^{8-n}:[1n],\dots [(n-1)n]
\end{equation}

For the $V_{4}^4$ there are 28 paths. Looking at the path $d^6u^2$

$$
\rightarrow_{d^{6}}\frac{(\bar\lambda_{L}\bar\lambda_{R})}{(4ik_+)^2}\nabla_{+}^{7}\delta(\lambda^{7}_{+})\nabla_{+}^{8}\delta(\lambda^{8}_{+})\Phi(y,\theta_{-})\rightarrow_{u}\frac{(\bar\lambda_{L}\sigma^{ij}\bar\lambda_{R})}{4(4ik_+)^2}(\nabla_{+}\sigma_{ij}\lambda_{+})\frac{\nabla_{+}^{7}}{\lambda_{+}^{7}}\nabla_{+}^{8}\delta(\lambda^{8}_{+})\Phi(y,\theta_{-})
$$

\begin{equation}
=\frac{(\bar\lambda_{L}\sigma^{ij}\bar\lambda_{R})}{4(4ik_+)^2}(\nabla_{+}\sigma_{ij})^{7}\nabla_{+}^{7}\nabla_{+}^{8}\delta(\lambda^{8}_{+})\Phi(y,\theta_{-})+\text{poles}
\end{equation}

Again, we drop the poles which will be treated in the next section. Proceeding with the last $u$ we have

\begin{equation}
\rightarrow_{u} \frac{(\bar\lambda_{L}\sigma^{ij}\bar\lambda_{R})(\bar\lambda_{L}\sigma^{kl}\bar\lambda_{R})}{(16ik_+)^2(\bar\lambda_L\bar\lambda_R)}(\nabla_{+}\sigma_{ij})^{8}(\nabla_{+}\sigma_{kl})^{7}\nabla_{+}^{7}\nabla_{+}^{8}\Phi(y,\theta_{-})+\text{poles} 
\end{equation}

The $d^{6}u^2$ path is the only one that produces a term with just two $\nabla_+$ of the form $\nabla^{8}_{+}\nabla_{+}^{7}$, so by $SO(8)$ symmetry we have

\begin{equation}
V_4^4=\frac{1}{2!}\frac{(\bar\lambda_{L}\sigma^{ij}\bar\lambda_{R})(\bar\lambda_{L}\sigma^{kl}\bar\lambda_{R})}{(16ik_+)^2(\bar\lambda_L\bar\lambda_R)}\frac{1}{2^2}(\nabla_{+}\sigma_{ij}\nabla_+)(\nabla_{+}\sigma_{kl}\nabla_+)\Phi(y,\theta_{-})
\end{equation}

In order to get rid of $(\bar\lambda_{L}\bar\lambda_{R})$ in the denominator we are going to need the following $SO(8)$ identities:

\begin{equation}
\sigma^{ij}_{[ab}\sigma^{kl}_{cd]}=\sigma^{[ij}_{[ab}\sigma^{kl]}_{cd]}+\frac{1}{2}\delta^{[i}_{[k}(\sigma^{j]m})_{[ab|}(\sigma_{l]m})_{|cd]}   
\end{equation}

and

\begin{equation}
\sigma^{[ij}_{\dot a\dot b}\sigma^{kl]}_{\dot c\dot d}=\sigma^{[ij}_{[\dot a\dot b}\sigma^{kl]}_{\dot c\dot d]}+\frac{1}{2}\delta^{[\dot a}_{[\dot c}(\sigma^{ijkl})^{\dot b]}_{\dot d]}  
\end{equation}

This two identities are related by triality, i.e. by rotating the indices $i\rightarrow a\rightarrow \dot a\rightarrow i$. 

The identity (44) implies that

\begin{equation}
(\nabla_+\sigma^{ij}\nabla_+)(\nabla_{+}\sigma^{kl}\nabla_+)=(\nabla_+\sigma^{[ij}\nabla_+)(\nabla_{+}\sigma^{kl]}\nabla_+) + \delta^{[i}_{[k}(\nabla_+\sigma^{j]m}\nabla_+)(\nabla_{+}\sigma_{l]m}\nabla_+)
\end{equation}

and since $(\sigma^{m}\bar\lambda_{L/R})^{a}(\sigma^{m}\bar\lambda_{L/R})^{b}=0$ due to the pure spinor constraint, and $\sigma^{ij}_{[ab}\sigma^{ij}_{cd]}=0$, just the part proportional to $(\nabla_{+}\bar\sigma^{[ij}\nabla_{+})(\nabla_{+}\bar\sigma^{kl]}\nabla_{+})$ contribute. Now, using the identity (45) and that $(\bar\lambda_{L/R}\bar\lambda_{L/R})=0$ we have

\begin{equation}
(\bar\lambda_{L}\bar\sigma^{[ij}\bar\lambda_{R})(\bar\lambda_{L}\bar\sigma^{kl]}\bar\lambda_{R})=(\bar\lambda_{L}\bar\lambda_{R})(\bar\lambda_{L}\sigma^{ijkl}\bar\lambda_{R})    
\end{equation}

so we can write

\begin{equation}
V_4^4=\frac{(\bar\lambda_{L}\sigma^{ijkl}\bar\lambda_{R})}{2!(32ik_+)^2}(\nabla_{+}\sigma_{ij}\nabla_+)(\nabla_{+}\sigma_{kl}\nabla_+)\Phi(y,\theta_{-})
\end{equation}

which again agrees with \cite{Berkovits:2019rwq}.

\subsection{Explicit cancellations for a few cases}

Instead of going all the way with a given path, in order to appreciate the cancellations of poles, it will be more convenient to compute each term $V_{2n}^{m}$ of the diagram. Note that all the elements of the diagonal that contribute with $V_{0}^0$ does not contain poles of $\lambda_{+}^a$. The $n$-th term of this diagonal is given by

\begin{equation}
V_{0}^{n}=(\bar\lambda_L\bar\lambda_R)\nabla^{9-n}_+\delta(\lambda_{+}^{9-n})\dots\nabla^{8}_+\delta(\lambda^{8}_+)\frac{\Phi(y,\theta_-)}{(4ik_+)^{n}}
\end{equation}

It is instructive to compute the term $V_{\,2}^{8}$ that is given by a unique path $du$.

$$
\rightarrow_{du}\left(\frac{(\bar\lambda_{L}\bar\sigma^{ij}\bar\lambda_R)}{4(\bar\lambda_{L}\bar\lambda_R)}\right)(\bar\lambda_L\bar\lambda_R)(\nabla_{+}\sigma_{ij}\lambda_{+})\left(\frac{\nabla^{2}_+}{\lambda_{+}^2}\right)\nabla^{3}_+\delta(\lambda^{3}_+)\dots\nabla^{8}_+\delta(\lambda^{8}_+)\frac{\Phi(y,\theta_{-})}{(4ik_+)^7}
$$

\begin{equation}
V_{2}^{8}=\left(\frac{(\bar\lambda_{L}\bar\sigma^{ij}\bar\lambda_R)}{16ik_+}\right)\sigma_{ij}^{12}\nabla^{1}_{+}\nabla^{2}_{+}\nabla^{3}_+\delta(\lambda^{3}_+)\dots\nabla^{8}_+\delta(\lambda^{8}_+)\frac{\Phi(y,\theta_-)}{(4ik_+)^6}
\end{equation}

It does not have poles. Now we can use $V_{2}^{8}$ and $V_{0}^{6}$ to compute $V_{2}^{7}$, which will be the sum of the following terms

\begin{equation}
V_{2}^{8}\rightarrow_{d}\left(\frac{(\bar\lambda_{L}\bar\sigma^{ij}\bar\lambda_R)}{16ik_+}\right)\sigma_{ij}^{12}\frac{3\lambda^{[1}_{+}\nabla^{2}_{+}\nabla^{3]}}{\lambda_+^{3}}\nabla^{4}_+\delta(\lambda^{4}_+)\dots\nabla^{8}_+\delta(\lambda^{8}_+)\frac{\Phi(y,\theta_-)}{(4ik_+)^{5}}
\end{equation}

$$
V_{0}^{6}\rightarrow_{u}\left(\frac{(\bar\lambda_{L}\bar\sigma^{ij}\bar\lambda_R)}{16ik_+}\right)(\nabla_{+}\sigma_{ij}\lambda_{+})\left(\frac{\nabla^{3}_+}{\lambda_{+}^3}\right)\nabla^{4}_+\delta(\lambda^{4}_+)\dots\nabla^{8}_+\delta(\lambda^{8}_+)\frac{\Phi(y,\theta_-)}{(4ik_+)^{5}}=
$$

\begin{equation}
=\left(\frac{(\bar\lambda_{L}\bar\sigma^{ij}\bar\lambda_R)}{16ik_+}\right)\left(\frac{2\sigma_{ij}^{12}\nabla^{[1}_+\lambda_{+}^{2]}\nabla^{3}_+}{\lambda_{+}^3}+\sigma_{ij}^{13}\nabla^{1}_+\nabla^{3}_++\sigma_{ij}^{23}\nabla^{2}_+\nabla^{3}_+\right)\dots\frac{\Phi(y,\theta_-)}{(4ik_+)^{5}}
\end{equation}

where the antisymmetrization of indices is of strength one. The sum of this two terms cancel the poles in $\lambda^{3}_+$ and gives

\begin{equation}
V_{2}^{7}=\left(\frac{(\bar\lambda_{L}\bar\sigma^{ij}\bar\lambda_R)}{16ik_+}\right)\left(\sigma_{ij}^{12}\nabla^{1}_+\nabla^{2}_++\sigma_{ij}^{13}\nabla^{1}_+\nabla^{3}_++\sigma_{ij}^{23}\nabla^{2}_+\nabla^{3}_+\right)\nabla^{4}_+\delta(\lambda^{4}_+)\dots\nabla^{8}_+\delta(\lambda^{8}_+)
\end{equation}

With $V_{2}^{7}$ we can obtain $V_{4}^{8}$ by a unique arrow, the arrow $u$. Note that now there is no $d$ arrow to cancel the the poles coming from the $u$ arrow, so the poles in $u$ should cancel by themselves.

$$
V_{2}^{7}\rightarrow_{u}\left(\frac{(\bar\lambda_{L}\bar\sigma^{ij}\bar\lambda_R)(\bar\lambda_{L}\bar\sigma^{kl}\bar\lambda_R)}{(16ik_+)^2(\bar\lambda_{L}\bar\lambda_R)}\right)\left(\sigma_{ij}^{12}\nabla^{1}_+\nabla^{2}_++\sigma_{ij}^{13}\nabla^{1}_+\nabla^{3}_++\sigma_{ij}^{23}\nabla^{2}_+\nabla^{3}_+\right)\times
$$

$$
\times (\nabla_{+}\sigma^{kl}\lambda_+)\left(\frac{\nabla^{4}_+}{\lambda^{4}_+}\right)\dots\nabla^{8}_+\delta(\lambda^{8}_+)\frac{\Phi(y,\theta_-)}{(4ik_+)^4}=
$$

$$
=\left(\frac{(\bar\lambda_{L}\bar\sigma^{ij}\bar\lambda_R)(\bar\lambda_{L}\bar\sigma^{kl}\bar\lambda_R)}{(16ik_+)^2(\bar\lambda_{L}\bar\lambda_R)}\right)\left[\left(\frac{\sigma_{ij}^{12}\nabla^{1}_+\nabla^{2}_+\nabla_{+}^{3}(\sigma_{kl}^{31}\lambda^{1}_++\sigma_{kl}^{32}\lambda^{2}_++\sigma_{kl}^{34}\lambda^{4}_+)}{\lambda^{4}_{+}}\right)\right.
$$

$$
+\left.\left(\frac{\sigma_{ij}^{13}\nabla^{1}_+\nabla^{3}_+\nabla_{+}^{2}(\sigma_{kl}^{21}\lambda^{1}_++\sigma_{kl}^{23}\lambda^{3}_++\sigma_{kl}^{24}\lambda^{4}_+)}{\lambda^{4}_{+}}\right)+\left(\frac{\sigma_{ij}^{23}\nabla^{2}_+\nabla^{3}_+\nabla_{+}^{1}(\sigma_{kl}^{12}\lambda^{2}_++\sigma_{kl}^{13}\lambda^{3}_++\sigma_{kl}^{14}\lambda^{4}_+)}{\lambda^{4}_{+}}\right)\right] 
$$

$$
\times \nabla^{4}_+\nabla^{5}_{+}\delta(\lambda^{5}_{+})\dots\nabla^{8}_+\delta(\lambda^{8}_+)\frac{\Phi(y,\theta_{-})}{(4ik_+)^4}
$$

Note that indeed the poles coming from $u$ cancel by themselves. The terms with $\lambda^{4}_{+}$ in the denominator cancels by the antisymmetry of $123$ and symmetry of $ij\leftrightarrow kl$, which implies that there is an anti-symmetry of the form $\lambda^{[a}_{+}\nabla^{b}_{+}\nabla^{c}_{+}\nabla^{d]}_{+}$. The result is

\begin{equation}
V_{4}^{8}=\frac{(\bar\lambda_{L}\sigma^{ijkl}\bar\lambda_{R})}{2!(32ik_+)^{2}}(\nabla_{+}\sigma^{ij}\nabla_{+})(\nabla_{+}\sigma^{kl}\nabla_{+})\nabla^{5}_{+}\delta(\lambda^{5}_{+})\dots\nabla^{8}_+\delta(\lambda^{8}_+)\frac{\Phi(y,\theta_-)}{(4ik_+)^4}
\end{equation}

\subsection{Computation of general $V_{2n}^{m}$}

One can expect that these cancellations of poles will also occur in the computation of a generic $V_{2n}^m$. In fact this can be proved. The statement we want to prove is the following: all the terms $V_{2n}^{m}$ of the diagram does not have poles in $\lambda_{+}^a=0$ and it is given by

\begin{equation}
V_{2n}^{m}=\frac{(\bar\lambda_{L}\bar\lambda_{R})}{n!(32ik_+)^n}\left(\frac{(\bar\lambda_{L}\sigma^{ij}\bar\lambda_{R})(\nabla_+\sigma^{ij}\nabla_+)}{(\bar\lambda_L\bar\lambda_R)}\right)^{n}\nabla^{9-m+2n}_+\delta(\lambda^{9-m+2n}_+)\dots\frac{\Phi(y,\theta_-)}{(4ik_+)^{m-2n}}
\end{equation}

We can prove this statement by induction. Assuming that $V_{2(n-1)}^{m-1}$ and $V_{2n}^{m+1}$ are of this particular form then $V_{2n}^m$ will be given by the sum of the two paths

$$
V_{2n}^{m+1}\rightarrow_{d}\left[\frac{8ink_+}{n!(32ik_+)^n}(\bar\lambda_{L}\sigma^{ij}\bar\lambda_{R})(\lambda_+\sigma^{ij}\nabla_{+})\left(\frac{(\bar\lambda_{L}\sigma^{ij}\bar\lambda_{R})(\nabla_+\sigma^{ij}\nabla_{+})}{(\bar\lambda_L\bar\lambda_R)}\right)^{n-1}\right.
$$

$$
\left.\times\frac{\nabla^{9-m-1+2n}_+}{\lambda^{9-m+2n}_+}+\frac{(\bar\lambda_{L}\bar\lambda_{R})}{n!(32ik_+)^n}\left(\frac{(\bar\lambda_{+}\sigma^{ij}\bar\lambda_{-})(\nabla_+\sigma^{ij}\nabla_+)}{(\bar\lambda_L\bar\lambda_R)}\right)^{n}\right]
$$

$$
\times\nabla^{9-m+2n}_+\delta(\lambda^{9-m+2n}_+)\dots\nabla^{8}_+\delta(\lambda^{8}_+)\frac{\Phi(y,\theta_-)}{(4ik_+)^{m-2n}}
$$

and

$$
V_{2(n-1)}^{m-1}\rightarrow_{u}\frac{1}{(n-1)!(32ik_+)^{n-1}}\frac{(\bar\lambda_{L}\sigma^{ij}\bar\lambda_{R})}{4}(\nabla_+\sigma^{ij}\lambda_+)\left(\frac{(\bar\lambda_{L}\sigma^{ij}\bar\lambda_{R})(\nabla_+\sigma^{ij}\nabla+)}{(\bar\lambda_L\bar\lambda_R)}\right)^{n-1}
$$

\begin{equation}
\times\frac{\nabla^{9-m-1+2n}_+}{\lambda^{9-m+2n}_+}\nabla^{9-m+2n}\delta(\lambda^{9-m+2n})\dots\nabla^{8}\delta(\lambda^{8})\frac{\Phi(y,\theta_-)}{(4ik_+)^{m-2n}}
\end{equation}

There is a cancellation between the first term coming from $V_{2n}^{m+1}$ and the whole expression coming from $V_{2(n-1)}^{m+1}$. This cancellation involves some finite part together with all the poles in $\lambda_{+}^a=0$. The result of this cancellation is precisely (55). For $V_{2n}^{8}$, the terms that are located at the top of the diagram, there is just contributions coming from the terms bellow, by the $u$ arrow, so there will be no cancellation between different paths. Let us compute this case:

$$
V_{2(n-1)}^{7}\rightarrow_{u}\frac{8ink_+}{n!(32ik_+)^{n}}(\bar\lambda_{L}\sigma^{ij}\bar\lambda_{R})(\nabla_{+}\sigma^{ij}\lambda_+)\left(\frac{(\bar\lambda_{L}\sigma^{ij}\bar\lambda_{R})(\nabla_+\sigma^{ij}\nabla_{+})}{(\bar\lambda_L\bar\lambda_R)}\right)^{n-1}
$$

\begin{equation}
\times\frac{\nabla^{1+2n}_+}{\lambda^{1+2n}_+}\nabla^{2+2n}_+\delta(\lambda^{2+2n}_+)\dots\nabla^{8}_+\delta(\lambda^{8}_+)\Phi(y,\theta_-)
\end{equation}

Since we have symmetry under $i_nj_n\leftrightarrow i_mj_m$ and $\sigma^{ij}_{ab}=-\sigma^{ij}_{ba}$, $V_{2n}^8$ is a sum of terms proportional to

\begin{equation}
\propto \lambda^{[a_1}_{+}\nabla^{a_2}_+\dots\nabla^{a_{2n-1}]}_+\frac{\nabla^{1+2n}_+}{\lambda^{1+2n}_+}\nabla^{2+2n}_+\delta(\lambda^{2+2n}_+)\dots\nabla^{8}_+\delta(\lambda^{8}_+)\Phi(y,\theta_{-})
\end{equation}

The only nonzero contribution is the one proportional to $\lambda^{2n+1}_{+}$. The terms proportional to $\lambda^{a}_+$ with $a$ lower than 1+2n will contain at least one $(\nabla^{a}_{+})^2$ which is zero and for $a$ greater than $1+2n$ it will be killed by the $\delta(\lambda_{+}^{a})$. The term that is left over is precisely the (55).

We proved that assuming the equation (55) is true for $V_{2(n-1)}^{m-1}$ and $V_{2n}^{m+1}$, when the last exist, then (55) is also true for $V_{2n}^{m}$. Since we already have shown that (55) hold for the first cases in the previous section the statement is now proved by induction.

Note that the presence of $(\bar\lambda_{L}\bar\lambda_{R})$ in the denominator can always disappears since there are products of $(\bar\lambda_{L}\sigma^{ij}\bar\lambda_{R})$ in the numerator. The $SO(8)$ identities (44) and (45), together with the pure spinor constraint, can be used to get rid of $(\bar \lambda_{L}\bar \lambda_{R})$ in the denominator. After that, the 0 picture vertex operator will be given by the sum $V=V_0+V_1+V_2+V_3+V_4$, where $V_{n}$ is given by

\begin{equation}
V_{n}=V_{2n}^{2n}= \frac{(\bar\lambda_{L}\bar\sigma^{i_1j_1\dots i_{n}j_{n}}\bar\lambda_{R})}{n!(32ik_+)^n}(\nabla_{+}\sigma^{i_1j_1}\nabla_{+})\dots(\nabla_{+}\sigma^{i_n j_n}\nabla_{+})\Phi(y,\theta_-)
\end{equation}

which is the same vertex operator of \cite{Berkovits:2019rwq}.

\section{Conclusion}

We have obtained a Type IIB vertex operator for the supergravity multiplet in terms of a light-cone complex superfield $\Phi(y,\theta_-)$ of \cite{Green:1983hw} \cite{Howe:1983sra} by following this new procedure proposed in \cite{Berkovits:2019ulm}\cite{Berkovits:2019rwq}. We also found that the Type IIB  massless vertex operators in the -8 picture takes a simple form

\begin{equation}
V_{-8}=\frac{(\bar \lambda_{L}\bar\lambda_{R})}{(4ik_+)^4}\bar\Phi(\bar y,\theta_+)\prod_{a}^{8}\delta (\lambda^{a}_+)
\end{equation}

It might be interesting to search for an amplitude prescription that works directly with this -8 picture vertex operators. Work is in progress to obtain such an amplitude prescription for both the flat background and the $AdS_{5}\times S^{5}$ background.

The computation of the zero picture vertex operator by picture raising the -8 picture vertex operator is more intricate in the $AdS_{5}\times S^{5}$ because of the $PSU(2,2|4)$ algebra. The -8 picture vertex operator for half-BPS state was constructed in \cite{Berkovits:2019rwq}\cite{Berkovits:2019ulm} and work is in progress to obtain the zero picture vertex operator from it.

For the Type IIA vertex operator, we can easily adapt what was done here by applying T duality along some direction, breaking the $SO(8)$ covariance. This will only change the definition of the variables $\theta_{\pm}^{a}$, $\lambda^{a}_{\pm}$ and $\bar\lambda_{\pm}^{\dot a}$, but will maintain the formulas of $V_{-8}$ and $V$ the same as before.

\section*{Acknowledgements}

$\quad$ I would like to thank Nathan Berkovits for giving me this project and clarifying various points, Dennis Zavaleta and Thiago Fleury for reading the manuscript and making important observations, and FAPESP grant 18/07834-8 for partial financial support.

\bibliography{ref}
\bibliographystyle{ieeetr}

\end{document}